\documentclass[11pt,twoside]{article}
\usepackage{asp2010}
\usepackage{graphicx}

\resetcounters

\begin{document}

\title{Binary population synthesis and SNIa rates}
\author{S. Toonen,$^1$ G. Nelemans,$^{1,2}$ M. Bours,$^{1,3}$ and S. Portegies Zwart$^4$
\affil{$^1$Department of Astrophysics, IMAPP, Radboud University Nijmegen, P.O. Box 9010, 6500 GL Nijmegen, The Netherlands}
\affil{$^2$Instituut voor Sterrenkunde, KU Leuven, Celestijnenstraat 200D, 3001 Leuven, Belgium}
\affil{$^3$Department of Physics, University of Warwick, Coventry CV4 7AL, United Kingdom}
\affil{$^4$Leiden Observatory, Leiden University,  P.O. Box 9513, 2300 RALeiden, The Netherlands}
}

\begin{abstract}
Despite the significance of type Ia supernovae (SNeIa) in many fields in astrophysics, SNeIa lack a theoretical explanation. We investigate the potential contribution to the SNeIa rate from the most common progenitor channels using the binary population synthesis (BPS) code SeBa. Using SeBa, we aim constrain binary processes such as the common envelope phase and the efficiency of mass retention of white dwarf accretion. We find that the simulated rates are not sufficient to explain the observed rates. Further, we find that the mass retention efficiency of white dwarf accretion significantly influences the rates, but does not explain all the differences between simulated rates from different BPS codes. 





\end{abstract}

\section{Introduction}

SNeIa are generally thought to be thermonuclear explosions of carbon/oxygen (CO) white dwarfs (WDs). The standard scenarios involve white dwarfs reaching the Chandrasekhar mass \citep[e.g.][]{Nom82}; either by hydrogen accretion from a non-degenerate companion \citep[single-degenerate channel, SD;][]{Whe73} or by a merger of two CO WDs \citep[double-degenerate channel, DD;][]{Web84, Ibe84}.
    
We investigate the contribution from the DD and SD channel to the SNIa rate with the binary population synthesis (BPS) codes SeBa \citep[][and Toonen et al. submitted]{Por96, Nel01}. BPS codes are very useful tools to study the evolution of binary stars and the processes that govern them. We study the SNIa delay time distribution (DTD), where the delay time is the time between the formation of the binary system and the SNIa event. In a simulation of a single burst of star formation the DTD gives the SNIa rate as a function of time after the starburst. The DTD is linked to the nuclear timescales of the progenitors and the binary evolution timescales up to the merger.

\section{Double degenerate channel}
\label{sec:dd}

We set out to predict SNIa rates for the double degenerate channel, with the additional constraint that our model corresponds well to the observed population of close double WDs (cDWDs) - of all flavours and masses. Even though there are no certain DD SNIa progenitors among the observed cDWDs, the DD SNIa and observed cDWD progenitors have gone through similar evolution paths and are affected by the same binary and stellar processes.  

    \begin{figure*}
    \centering
    \begin{tabular}{c c}
	\includegraphics[scale=0.3, angle=270]{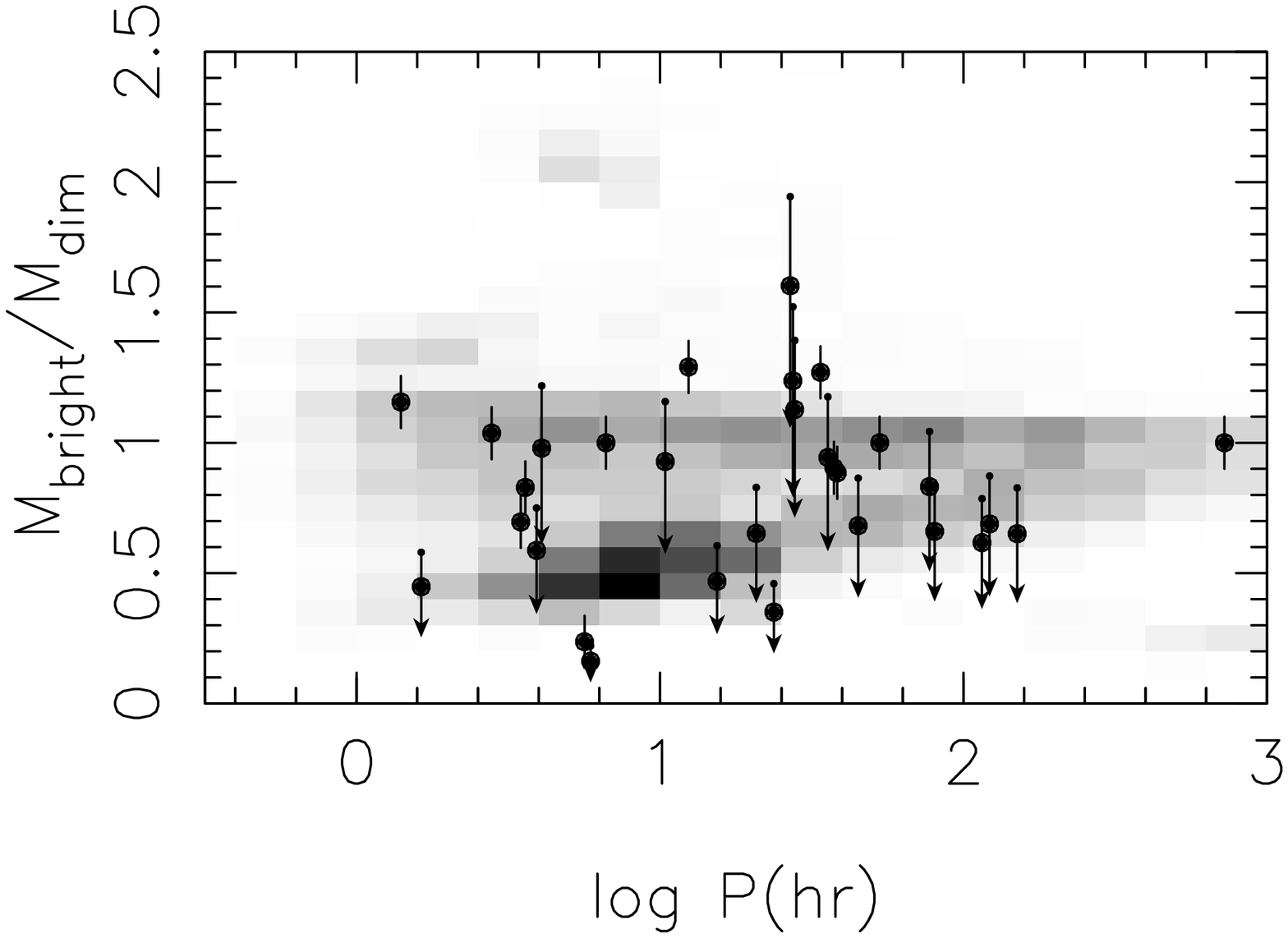} &
	\includegraphics[scale=0.3, angle=270]{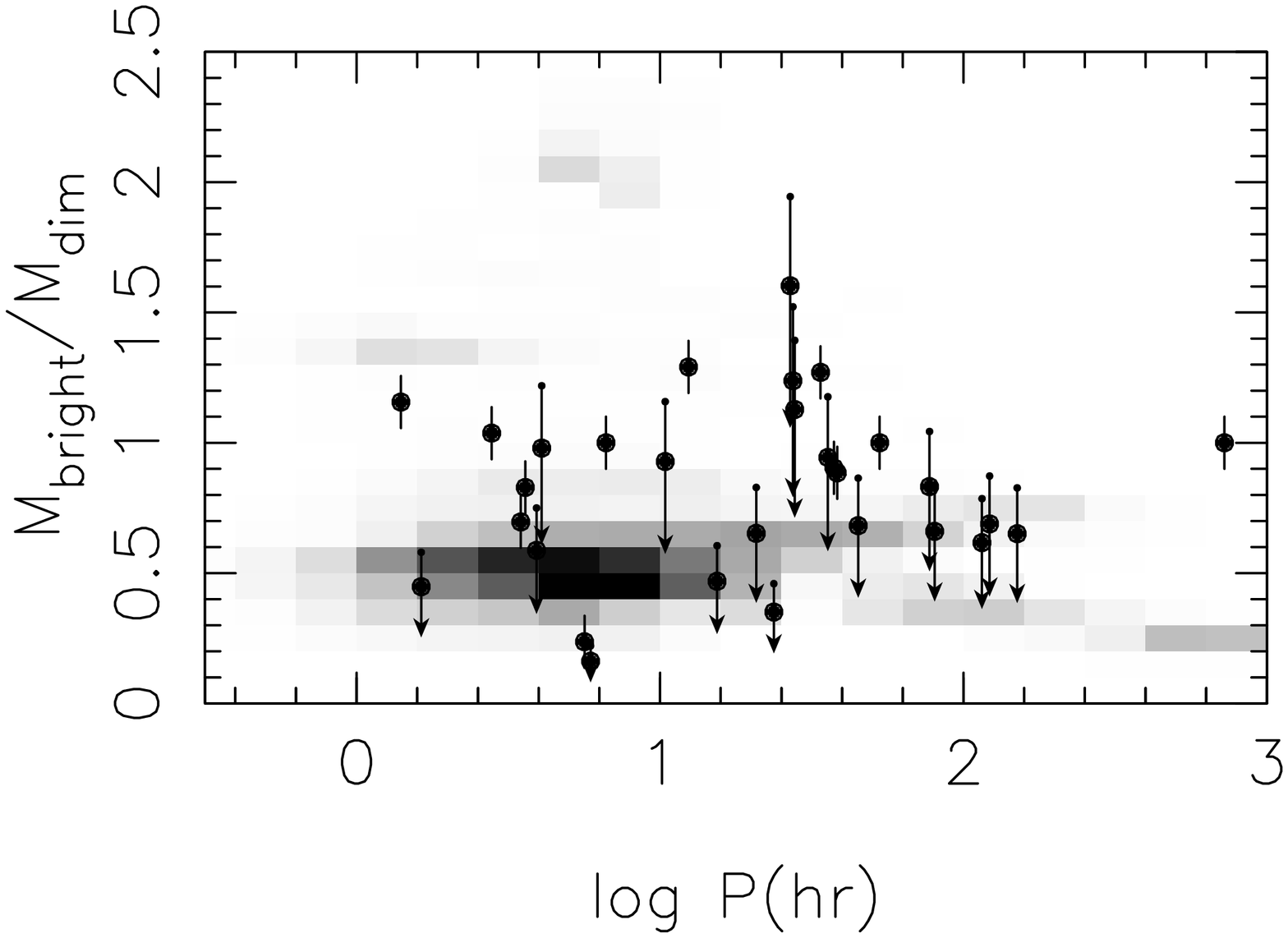} \\
	(a) & (b) \\
	\end{tabular}
    \caption{Simulated distribution of population of visible double white dwarfs  as a function of orbital period and mass ratio, where mass ratio is defined as the mass of the brighter white dwarf divided by that of the dimmer white dwarf. Left model $\gamma \alpha$ is used, on the right model $\alpha \alpha$. The intensity of the grey scale corresponds to the density of objects on a linear scale. The same grey scale is used for both plots. Observed binary white dwarfs \citep{Mar11} are overplotted with filled circles. 
    } 
    \label{fig:pop_dwd}
    \end{figure*}

cDWDs are believed to encounter at least two phases of common envelope (CE) evolution. In spite of the importance of the CE phase, it remains poorly understood. Several prescriptions for CE evolution have been proposed. The $\alpha$ formalism (\citet{Web84}) is based on the conservation of orbital energy and the $\gamma$ formalism  (\citet{Nel00}) is based on the conservation of angular momentum. In model $\alpha\alpha$ the $\alpha$ formalism is used to determine the outcome of every CE. For model $\gamma\alpha$ the $\gamma$-prescription is applied unless the binary contains a compact object or the CE is triggered by the Darwin-Riemann instability \citep{Dar1879}. The $\gamma\alpha$-model reproduces the mass ratio distribution of the observed double white dwarfs best, see Fig. \ref{fig:pop_dwd}.

The DTD of model $\gamma\alpha$ and model $\alpha\alpha$ are similar in showing strong declines with time and comparable time-integrated rates of $2.0\cdot 10^{-4}\ M_{\odot}^{-1}$ resp. $3.3 \cdot 10^{-4}\ M_{\odot}^{-1}$. Most importantly, the simulated time-integrated numbers do not match the observed number of $2.3\pm 0.6 \cdot 10^{-3}\ M_{\odot}^{-1}$ \citep{Mao11} by a factor of $\sim 7-12$. 
Many things influence the normalisation of the SNIa rates; the assumed binary fraction, metallicity, initial distribution of masses and orbital parameters. However, preliminary results show that the integrated rates are not affected by factors sufficient to match the observed rate. If so, the main contribution to the SNIa rate comes from other channels as for example the single degenerate scenario (e.g. supersoft sources), double detonating sub-Chandrasekhar accretors \citep[see e.g.][]{Kro10} or Kozai oscillations in triple systems (\citet{Sha12}; Hamers et al. in prep.). For more information about our study, see Toonen, Nelemans, Portegies Zwart submitted. 


\begin{figure*}
\centering
\includegraphics[scale = 0.3]{{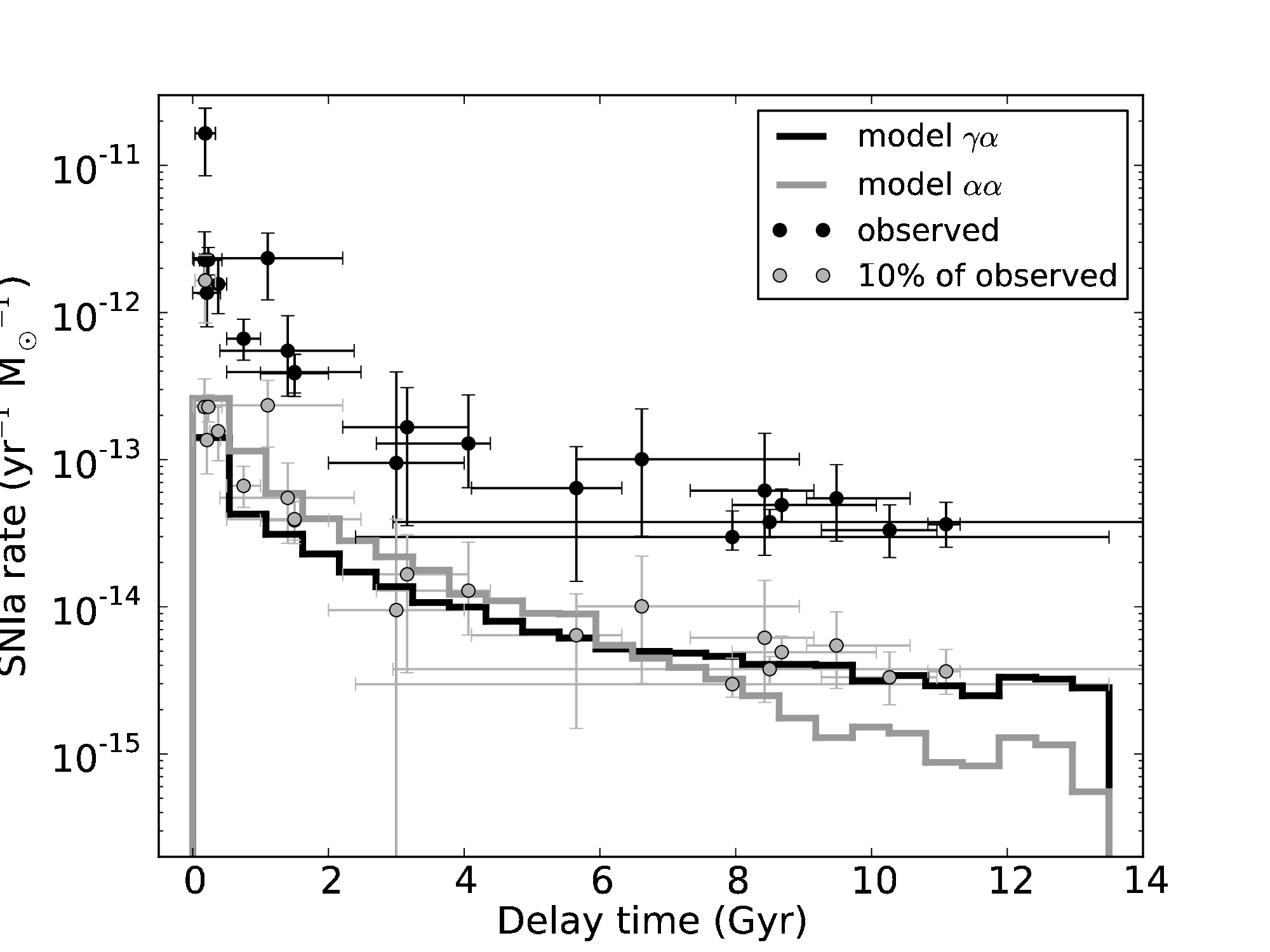}}
\caption{SNIa rate in the DD channel per yr per M$_{\odot}$ formed stellar mass as a function of delay time. 
Delay times are shown for two different prescriptions of the CE phase; model $\gamma\alpha$ and model $\alpha\alpha$, see section \ref{sec:dd}. Overplotted with black circles are the observed values of the SNIa rate \citep[see][for a review]{Mao11b}.
For comparison the grey circles show the observations scaled down by a factor 10.}
\label{fig:DD}
\end{figure*}

\section{Single degenerate channel}
\begin{figure*}
    \centering
    \begin{tabular}{c c}
	\includegraphics[scale = 0.23]{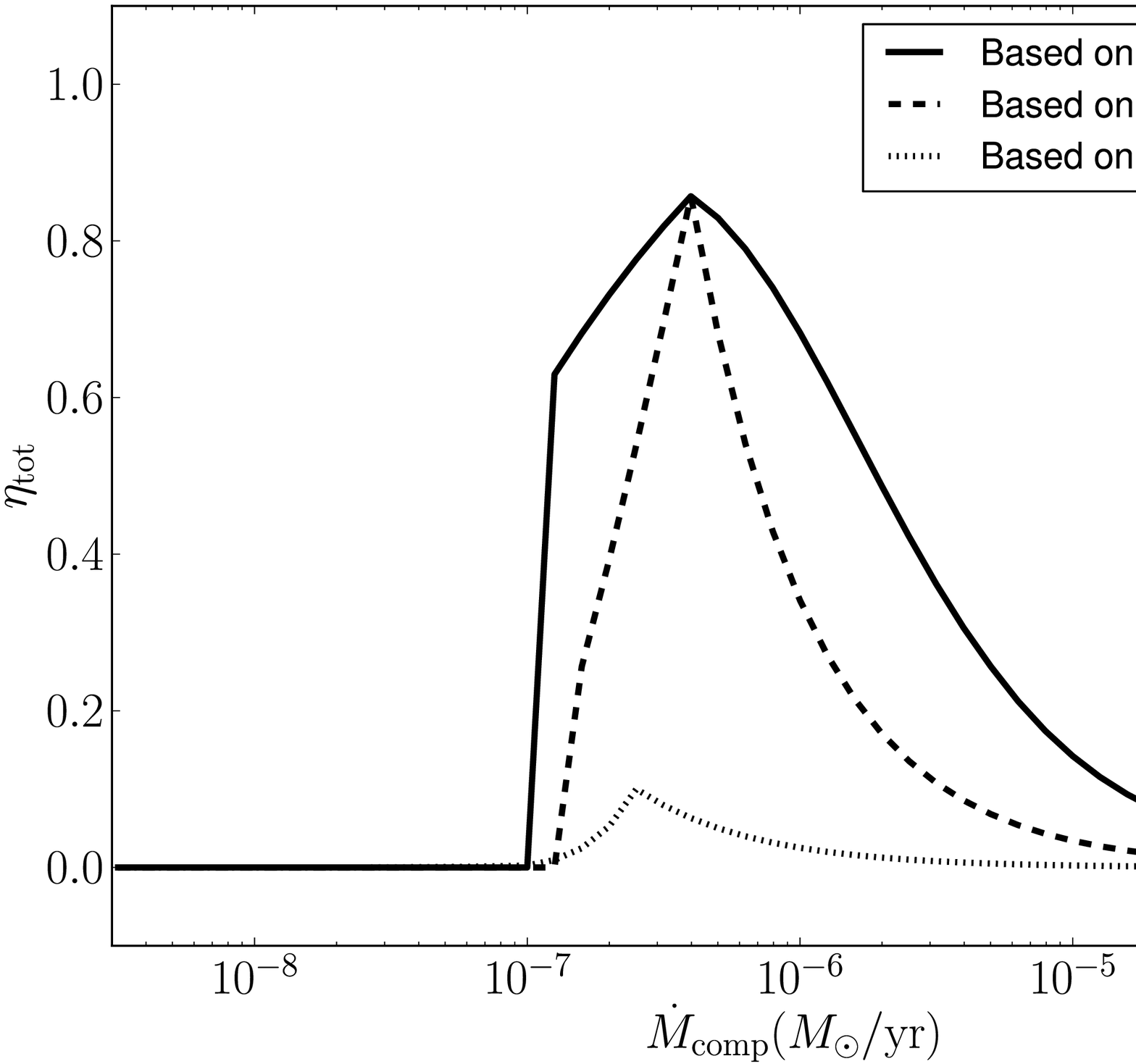} &
	\includegraphics[scale = 0.3]{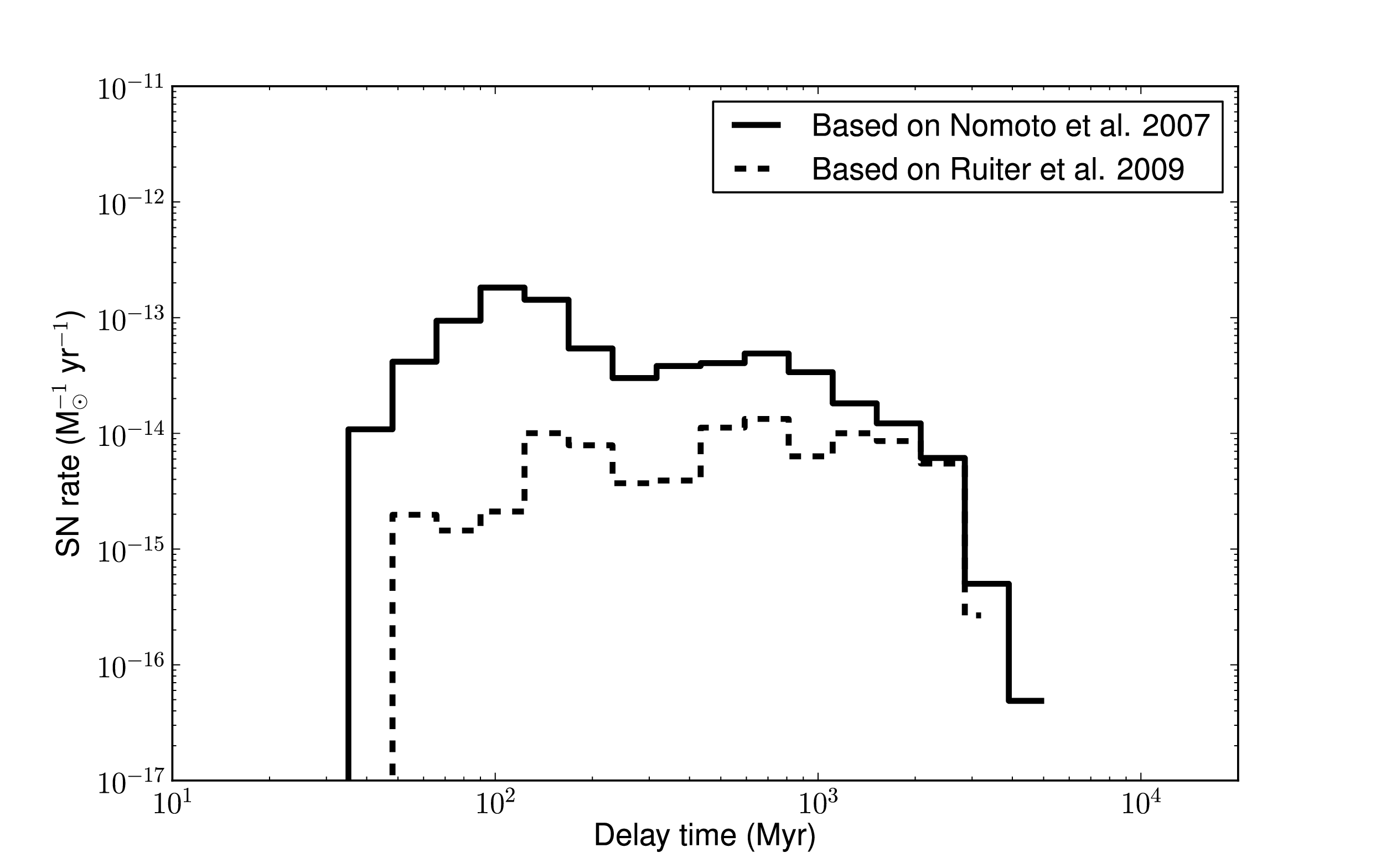} \\
	a) & b) \\
	\end{tabular}
    \caption{a) The mass retention efficiency of WD hydrogen accretion as a function of the mass transfer rate. $\eta_{\rm tot}$ represents the fraction of the transferred mass that is retained by the WD. If $\eta_{\rm tot}=0$ no mass is retained by the WD, when $\eta_{\rm tot}=1$ the WD accretes matter most effectively. In the figure we have assumed the white dwarf to be of one solar mass. b) The delay time distribution that result from the different retention efficiencies discussed in the text. Note that we found no SNIa when the retention efficiency of \citet{Yun10} are used.} 
    \label{fig:SD}
    \end{figure*}

Theoretical rates for the SD channel predicted by different BPS codes vary over four orders of magnitude and do not agree with each other or with observational data \citep{Nel12}. The exact origin of these differences remains unclear. We therefore study the effect of the efficiency of WD accretion, which is poorly understood because of processes such as novae and stable burning. Three prescriptions are used \citep{Nom07, Rui09, Yun10} that are based on \citet{Nom07} and \citet{Pri95} in different combinations which differ strongly, see Fig. \ref{fig:SD}a. We use the BPS code SeBa to simulate the SNIa rates in the SD channel for each retention efficiency. The simulated SNIa rates are significantly affected by the prescription used for the mass retention efficiency. The integrated rates vary between $7.0\cdot 10^{-5}\ M_{\odot}^{-1}$,  $2.2 \cdot 10^{-5}\ M_{\odot}^{-1}$ and an upper limit of $1 \cdot 10^{-7}\ M_{\odot}^{-1}$ when the retention efficiency is used as in \citet{Nom07}, \citet{Rui09} resp. \citet{Yun10} where the observed rate is $2.3\pm 0.6 \cdot 10^{-3}\ M_{\odot}^{-1}$ \citep{Mao11}. However, the retention efficiency does not explain all differences between the theoretical SNIa rate distributions from different codes. For more information about this study, see Bours, Toonen, Nelemans in prep.

\section{Outlook}
To understand the differences in the predictions of the various BPS codes in the SD and DD scenario, we started a collaboration to compare three BPS codes. The codes involved are the Binary\_c code \citep{Izz06, Cla11}, the Brussels code \citep{DeD04, Men10} and SeBa. The comparison focuses on the evolution of low and intermediate mass binaries containing one or more white dwarfs. The goal is to investigate whether differences in the simulated populations are due to numerical effects, or whether they can be explained by differences in the input physics. The comparison indicates that the differences are caused by varying assumptions for the input physics. For more information about this study, see the contribution to this volume by Claeys et al.

\bibliographystyle{asp2010}
\bibliography{bibtex_silvia_toonen}

\end{document}